\documentclass[fleqn,twoside]{article}
\usepackage{espcrc2}

\usepackage{graphicx}

\DeclareMathAlphabet   {\mathsc}{OT1}{cmr}{m}{sc}

\newcommand{\AmS}{{\protect\the\textfont2
  A\kern-.1667em\lower.5ex\hbox{M}\kern-.125emS}}
\def\[{\left [}
\def\]{\right ]}
\def\({\left (}
\def\){\right )}
\newcommand{\beq}{\begin{equation}}
\newcommand{\eeq}{\end{equation}}
\newcommand{\bea}{\begin{eqnarray}}
\newcommand{\eea}{\end{eqnarray}}

\newcommand{\lbr}{\left\{}
\newcommand{\rbr}{\right\}}
\newcommand{\order}{\mathcal{O}}

\newcommand{\GeV}{~\mathrm{GeV}}

\newcommand{\EW}{\mathsc{ew}}

\newcommand{\DSaa}{(\Delta S_{AA})^2}
\newcommand{\DSab}{(\Delta S_{AB})^2}
\newcommand{\met}{\not{\hspace{-.05in}{E_T}}}
\newcommand{\SUSY}{\mathsc{susy}}
\newcommand{\SM}{\mathsc{sm}}
\newcommand{\slashed}[1]{\not{\hspace{-.05in}#1}}
\newcommand{\wtd}[1]{\widetilde{#1}}
\newcommand{\gappeq}{\mathrel{\rlap {\raise.5ex\hbox{$>$}}
{\lower.5ex\hbox{$\sim$}}}}
\newcommand{\lappeq}{\mathrel{\rlap{\raise.5ex\hbox{$<$}}
{\lower.5ex\hbox{$\sim$}}}}

\title{Testing Gaugino Mass Unification Directly at the LHC}

\author{Brent D. Nelson\address[MCSD]{Department of Physics,
        Northeastern University, Boston, Massachusetts, USA}%
        \thanks{Proceedings of talk given at
        the international workshop ``Beyond the Standard Model Physics and
        LHC Signatures'' (BSM-LHC). The work described was done in collaboration with
        B.~Altunkaynak, P.~Grajek, M.~Holmes and G.~Kane~\cite{Altunkaynak:2009tg}
        and supported by NSF grant PHY-0653587.}
        .}

\begin{document}

\begin{abstract}
We report on the first step of a systematic study of how gaugino
mass unification can be probed at the LHC in a quasi-model
independent manner. Here we focus our attention on the theoretically
well-motivated mirage pattern of gaugino masses, a one-parameter
family of models of which universal (high scale) gaugino masses are
a limiting case. Using a statistical method to optimize our
signature selection we arrive at three ensembles of observables
targeted at the physics of the gaugino sector, allowing for a
determination of this non-universality parameter without
reconstructing individual mass eigenvalues or the soft
supersymmetry-breaking gaugino masses themselves. In this controlled
environment we find that approximately 80\% of the supersymmetric
parameter space would give rise to a model for which our method will
detect non-universality in the gaugino mass sector at the 10\% level
with $\order(10\,{\rm fb^{-1}})$ of integrated luminosity.
\vspace{1pc}
\end{abstract}

\maketitle

\section{MOTIVATION}

Given that the LHC era is nearly here, it is hardly surprising to
find members of the high energy theory community are increasingly
turning their attention away from esoteric issues of model-building
and towards issues of directly confronting theory with data. This
activity has produced a number of new measurement techniques.
Previously unconsidered signatures have been developed by studying
certain ``what-if'' scenarios, and a new emphasis has been placed on
making concrete predictions from famously nebulous string
constructions. All of this represents important progress, but here I
wish to consider another aspect of the challenge that lies ahead.
Specifically, I wish to imagine the state of our field three to five
years from now. Assuming something like the minimal supersymmetric
standard model (MSSM) is discovered, as so many expect, then we can
imagine being presented with a number of measurements that are
indirectly related to the underlying soft supersymmetry-breaking
mechanisms. Here I will discuss what is likely just the first step
in a research program into what comes next: how to connect the
multiple LHC observations to organizing principles in some
(high-energy) effective Lagrangian of underlying physics.

In addressing this issue we must be careful to avoid the pitfalls of
the ``LHC Inverse'' problem: the empirical fact that even in very
restrictive model frameworks it is quite likely that more than one
set of model parameters will give predictions for LHC observations
that are in good agreement with the experimental
data~\cite{ArkaniHamed:2005px}. Much recent work has focused on how
to address this
issue~\cite{Kane:2006hd,Berger:2007yu,Berger:2007ut,Altunkaynak:2008ry,Kane:2008gb,Balazs:2009it},
and we will borrow much of the philosophy and many of the useful
techniques from this recent literature. The LHC inversion problem
generally results from trying to use a large ensemble of
(correlated) observations to constrain a multi-dimensional parameter
space. But our ultimate goal {\em as theorists} is to understand
broad properties of the {\em underlying physics itself}. Ever more
precise measurement of a single parameter -- say the gluino mass --
does not necessarily further this goal. The most important such
broad property is undoubtedly the issue of gaugino mass unification:
few properties of the superpartner spectrum have more far-reaching
implications for low-energy phenomenology, the nature of
supersymmetry breaking, and the structure of the underlying physics
Lagrangian~\cite{Binetruy:2005ez,Choi:2007ka}.
But these soft parameters are not themselves directly measurable at
the LHC~\cite{Kneur:1998gy}.
One might consider performing a fit to some particular theory, such
as minimal supergravity (mSUGRA), in which universal gaugino masses
are assumed~\cite{Arnowitt:1992aq}, but we are not so much
interested in whether mSUGRA -- or any other particular theory for
which gaugino mass universality is a feature -- is a good fit to the
data. Rather, we wish to know whether gaugino mass universality is a
property of the underlying physics {\em independent of all other
properties of the model.}

We will therefore begin our attack by considering a concrete
parametrization of non-universalities in soft gaugino masses. In
recent work by Choi and Nilles~\cite{Choi:2007ka} soft
supersymmetry-breaking gaugino mass patterns were explored in a
variety of string-motivated contexts. In particular, the so-called
``mirage pattern'' of gaugino
masses~\cite{Choi:2004sx,Choi:2005ge,Falkowski:2005ck} provides an
interesting case study in gaugino mass non-universality. This
paradigm gets its name from the fact that should the mirage pattern
of gaugino masses be used as the low-energy boundary condition of
the (one-loop) renormalization group equations then there will exist
some high energy scale at which all three gaugino masses are
identical~\cite{Choi:2005uz}. The set of all such low-energy
boundary conditions that satisfy the mirage condition defines a
one-parameter family of models. In the parametrization we adopt
from~\cite{Choi:2007ka}, the gaugino mass ratios at the electroweak
scale take the form
\begin{equation}
M_1:M_2:M_3\simeq (1+0.66\alpha):(2+0.2\alpha):(6-1.8\alpha)
\label{mirage_ratios}
\end{equation}
where the case $\alpha=0$ is precisely the unified mSUGRA limit. In
the limit of very large values for the parameter $\alpha$ the ratios
among the gaugino masses approach those of the anomaly-mediated
supersymmetry breaking (AMSB)
paradigm~\cite{Giudice:1998xp,Randall:1998uk}. In fact, the mirage
pattern is most naturally realized in scenarios in which a common
contribution to all gaugino masses is balanced against an equally
sizable contribution proportional to the beta-function coefficients
of the three Standard Model gauge groups. Such an outcome arises in
string-motivated contexts, such as KKLT-type moduli stabilization in
D-brane models~\cite{Kachru:2003aw,Grana:2005jc} and K\"ahler
stabilization in heterotic string
models~\cite{Gaillard:2007jr,Binetruy:1996xja,Binetruy:1996nx,Casas:1996zi,Binetruy:1997vr,Gaillard:1999et,Binetruy:2000md,Kane:2002qp}.
Importantly, however, it can arise in {\em non}-stringy models, such
as deflected anomaly mediation~\cite{Katz:1999uw,Rattazzi:1999qg}.
We note that in none of these cases is the pure-AMSB limit likely to
be obtained, so our focus here will be on small to moderate values
of the parameter $\alpha$ in~(\ref{mirage_ratios}).

\section{METHOD}

The ultimate goal is to ask whether or not soft supersymmetry
breaking gaugino masses obey some sort of universality condition
independent of all other facts about the supersymmetric model. We
begin by asking a simpler question: {\em assuming} the world is
defined by the MSSM with gaugino masses obeying the
relation~(\ref{mirage_ratios}), how well can we determine the value
of the parameter~$\alpha$. At the very least we would like to be
able to establish that $\alpha \neq 0$ with a relatively small
amount of integrated luminosity. The first step in such an
incremental approach is to demonstrate that some set of ``targeted
observables''~\cite{Binetruy:2003cy} (we will call them
``signatures'' in what follows) is sensitive to small changes in the
value of the parameter $\alpha$ in a world where all other
parameters which define the SUSY model are kept fixed.

\subsection{Setting Up the Problem}
\label{sec:setup}

We will construct what we will call a ``model line'' by specifying
the supersymmetric model in all aspects other than the gaugino
sector. We choose a simplified set of 17~input parameters given by
\beq \lbr \begin{array}{c} \tan\beta,\,\,m^2_{H_u},\,\,m^2_{H_d} \\
M_3,\,\,A_t,\,\,A_b,\,\,A_{\tau} \\
m_{Q_{1,2}},\,m_{U_{1,2}},\,m_{D_{1,2}},\,m_{L_{1,2}},\,m_{E_{1,2}} \\
m_{Q_3},\,m_{U_3},\,m_{D_3},\,m_{L_3},\,m_{E_3} \end{array} \rbr \,
, \label{paramset} \eeq
which are understood to be taken at the electroweak scale
(specifically $\Lambda_{\EW} = 1000\GeV$) so no renormalization
group evolution is required. The gluino soft mass $M_3$ will set the
overall scale for the gaugino mass sector. The other two gaugino
masses $M_1$ and $M_2$ are then determined relative to $M_3$
via~(\ref{mirage_ratios}). A model line will take the inputs
of~(\ref{paramset}) and then construct a family of theories by
varying the parameter $\alpha$ from $\alpha=0$ (the mSUGRA limit) to
some non-zero value in even increments.

For each point along the model line we passed the model parameters
to {\tt PYTHIA 6.4} for spectrum calculation and event generation.
Events are then sent to the {\tt PGS4} package to simulate the
detector response. Additional details of the analysis will be
presented in later sections. The end result of our procedure is a
set of observable quantities that have been designed and (at least
crudely) optimized so as to be effective at separating $\alpha = 0$
from other points along the model line in the least amount of
integrated luminosity possible.
%

\subsection{Distinguishability}
\label{sec:separate}

The technique we employ to distinguish between candidate theories
using LHC~observables involves the construction of a variable
similar to a traditional chi-square statistic
\beq (\Delta S_{AB})^2 = \frac{1}{n} \, \sum_i \[ \frac{S_i^A -
S_i^B}{\delta S_i^{AB}}\]^2, \label{DeltaS} \eeq
where $S$ is some observable quantity (or signature). The index
$i=1,\dots,n$ labels these signatures, with $n$ being the total
number of signatures considered. The labels $A$ and $B$ indicate two
distinct theories which give rise to the signature sets $S_i^A$ and
$S_i^B$, respectively. Finally, the error term $\delta S_i^{AB}$ is
an appropriately-constructed measure of the uncertainty of the term
in the numerator, i.e. the difference between the signatures.
%
In this work we will always define a signature $S$ as an observation
interpreted as a count (or number) and denote it with capital $N$.
One example is the number of same-sign, same-flavor lepton pairs in
a certain amount of integrated luminosity. Another example is taking
the invariant mass of all such pairs and forming a histogram of the
results, then integrating from some minimum value to some maximum
value to obtain a number. In principle there can be an infinite
number of signatures defined in this manner. In practice
experimentalists will consider a finite number and many such
signatures are redundant.

We can identify any signature $N_i$ with an effective cross section
$\bar{\sigma}_i$ via the relation
\beq \bar{\sigma}_i = N_i / L\, ,  \label{barsigma} \eeq
where $L$ is the integrated luminosity. We refer to this as an {\em
effective} cross-section as it is defined by the counting signature
$N_i$ which contains in its definition such things as the geometric
cuts that are performed on the data, the detector efficiencies, and
so forth. Furthermore these effective cross sections, whether
inferred from actual data or simulated data, are subject to
statistical fluctuations. The transformation in~(\ref{barsigma})
allows for a comparison of two signatures with differing amounts of
integrated luminosity. This will prove useful in cases where the
experimental data is presented after a limited amount of integrated
luminosity $L_A$, but the simulation being compared to the data
involves a much higher integrated luminosity $L_B$.
We will assume that the errors associated with the signatures $N_i$
are purely statistical in nature and that the integrated
luminosities $L_A$ and $L_B$ are precisely known, so that $(\Delta
S_{AB})^2$ is given by
\beq (\Delta S_{AB})^2 = \frac{1}{n} \, \sum_i \[
\frac{\bar{\sigma}_i^A - \bar{\sigma}_i^B}{\sqrt{ \bar{\sigma}_i^A /
L_A  + \bar{\sigma}_i^B / L_B }}\]^2, \label{DeltaS2} \eeq
where each cross section includes the (common) Standard Model
background, i.e. $\bar{\sigma}_i = \bar{\sigma}^{\SUSY}_i +
\bar{\sigma}^{\SM}$.

The variable $\DSab$ forms a measure of the distance between any two
theories in the space of signatures defined by the $S_i$. We can use
this metric on signature space to answer the following question: how
far apart should two sets of signatures $S_i^A$ and $S_i^B$ be
before we conclude that theories $A$ and $B$ are truly distinct? To
answer this question we note that in the limit in which the
luminosities $L_A$ and $L_B$ are large the probability distribution
for the quantity $\DSab$ given by
\begin{equation}
P(\Delta S^2) = n \, \chi_{n,\lambda}^2(n \Delta S^2)\, ,
\label{prob}
\end{equation}
where $\chi_{n,\lambda}^2$ is the non-central chi-squared
distribution for $n$ degrees of freedom. The non-centrality
parameter $\lambda$ is given by
\beq \lambda = \sum_i \frac{(\sigma_i^A - \sigma_i^B)^2}{\sigma_i^A
/ L_A  + \sigma_i^B / L_B }\, , \label{lambdadef} \eeq
and now the $\sigma_i$ represent {\em exact} cross sections.
From~(\ref{prob})~and~(\ref{lambdadef}) it is apparent that all the
physics behind the distribution of possible $\DSab$ values is
contained in the values of $\lambda$ and $n$.

The problem of defining ``distinguishability'' has now been reduced
to a well-understood problem in statistics which can be solved
analytically. To say that two potential models have been
distinguished -- or that a set of $n$ observations are inconsistent
with those derived from a simulation -- we first demand that the
quantity $\DSab$ constructed from the $n$ observations satisfy
$\DSab > \DSaa\big|_{\rm 95th}$. In other words, we first require
that the distance in signature space be larger than that expected
simply by quantum fluctuations with 95\% confidence. The values of
$\DSaa\big|_{\rm 95th}$ for $\lambda = 0$ are easily computed and
tabulated for any value of $n$.

\subsection{Optimization}
\label{sec:optimize}

Now return to the basic problem: using experimental data to
distinguish two models that truly are distinct. Though we have
$\lambda \neq 0$, there is always a finite chance that an
experimental measurement will not reveal this fact due to quantum
fluctuations. For any given value of $\lambda \neq 0$, the
probability that a measurement of $\DSab$ will fluctuate to a value
so small that it is not possible to separate two distinct models (to
confidence level $p$) is simply the fraction of the probability
distribution in~(\ref{prob}) that lies to the left of the value
$\DSaa\big|_{\rm 95th}$. There is always some minimum value of the
non-centrality parameter that can be chosen so that this fraction is
below some pre-determined value. Let us call that value
$\lambda_{\rm min}(n,p)$. Again, these values are simply determined
by integrating the probability distributions~(\ref{prob}) and are
independent of the physics of the problem. So, for example, given
two distinct models~$A$ and~$B$, any combination of $n$ experimental
signatures such that $\lambda > \lambda_{\rm min}(n,p=0.95)$ will be
effective in demonstrating that the two models are indeed different
95\% of the time, with a confidence level of 95\%.

Let us assume for the moment that ``model $A$'' is the experimental
data, which corresponds to an integrated luminosity of $L^{\rm
exp}$. Our ``model $B$'' can then be a simulation with integrated
luminosity $L^{\rm sim} = q L^{\rm exp}$. Define the quantity
\begin{equation} R_{AB} = \sum_i (R_{AB})_i = \sum_i
\frac{ (\sigma_i^A - \sigma_i^B)^2 }{ \sigma_i^A + \frac{1}{q} \,
\sigma_i^B} \, \label{Rdef} \end{equation}
where $R_{AB}$ has the units of a cross section. Our condition for
95\% certainty that we will be able to separate two truly distinct
models at the 95\% confidence level becomes
\begin{equation}  L_{\rm exp}  \geq \frac{\lambda_{\rm min}(n,0.95)}{R_{AB}}\,
. \label{ourcond} \end{equation}
Given two models~$A$ and~$B$ and a selection of $n$ signatures both
$\lambda_{\rm min}(n,0.95)$ and $R_{AB}$ are completely determined.
Therefore the minimum amount of integrated luminosity needed to
separate the models experimentally will be given by
\begin{equation}  L_{\rm min}(p) = \frac{\lambda_{\rm
min}(n,p)}{R_{AB}}\, . \label{Lmin} \end{equation}
Equation~(\ref{Lmin}) provides us with a quantitative measure of the
efficacy of any particular set of $n$ (uncorrelated) signatures. We
therefore wish to select a set of $n$ signatures $S_i$ such that the
quantity $L_{\rm min}(p)$ as defined in~(\ref{Lmin}), for a given
value of $p$, is as small as it can possibly be over the widest
possible array of model pairs~$A$ and~$B$. We must also do our best
to ensure that the $n$~signatures we choose to consider are
reasonably uncorrelated with one another so that the statistical
treatment of the preceding section is applicable.

The optimal strategy is generally to choose a rather small subset of
the total signatures one could imagine. In part this is because in
any set of observations the probability that any two are highly
correlated with one another will only grow with the size of the set.
But in addition not all possible signatures are equally effective at
separating models.
An absolute measure of the ``power'' of any given signature to
separate two models~$A$ and~$B$ can be provided by considering the
condition in~(\ref{Lmin}). For any signature $S_i$ we can define an
individual $(L_{\rm min})_i$ by
\begin{equation}
(L_{\rm min})_i = \lambda_{\rm min}(1,p) \, \frac{\sigma_i^A +
\frac{1}{q} \, \sigma_i^B}{(\sigma_i^A - \sigma_i^B)^2} \,
,\label{Lmindef}
\end{equation}
where, for example, $\lambda_{\rm min}(1,0.95) = 12.99$. This
quantity is exactly the integrated luminosity required to separate
models $A$ and $B$, to confidence level $p$, by using the single
observable $S_i$.
As more signatures are added, the threshold for adding the next
signature in the list gets steadily stronger. Mathematically, the
ratio  $\lambda_{\rm min}(n,p) / \lambda_{\rm min}(1,p)$ grows with
increasing $n$. This indicates that as we add signatures with ever
diminishing $(L_{\rm min})_i$ values we will eventually encounter a
point of negative returns, where the resulting overall $L_{\rm min}$
starts to grow again.

For a particular pair of models, $A$~and~$B$, it is always possible
to find the optimal list of signatures from among a given grand set
by ordering the resulting $(L_{\rm min})_i$ values and adding them
sequentially until a minimum of $L_{\rm min}$ is observed. To do so,
we note that kinematic distributions must be converted into counts
(and all counts are then converted into effective cross sections).
This conversion requires specifying an integration range for each
histogram. The choice of this range can itself be optimized, by
considering each integration range as a separate signature and
choosing the values such that $(L_{\rm min})_i$ is minimized. By
repeating this optimization procedure over many background model
lines it is possible to construct a list that will be at least {\em
close} to optimal over a wide range of supersymmetric model space.

\section{SIMULATION AND SIGNATURE SELECTION}
\label{sec:simulation}

\begin{table*}[t]
\caption{Signature List B. Distributions are integrated from ``Min
Value'' to ``Max Value''.} \label{tbl:sigB}
\newcommand{\m}{\hphantom{$-$}}
\newcommand{\cc}[1]{\multicolumn{1}{c}{#1}}
\renewcommand{\tabcolsep}{2pc} 
\renewcommand{\arraystretch}{1.2} 
\begin{tabular}{@{}llcc}
\hline
 & Description & Min Value & Max Value \\  \hline
%
%
1 & $M_{\rm eff}^{\rm jets}$ [0 leptons, $\geq 5$ jets] & 1100 GeV & End \\
2 & $M_{\rm eff}^{\rm any}$ [0 leptons, $\leq 4$ jets]
& 1450 GeV & End \\
3 & $M_{\rm eff}^{\rm any}$ [$\geq 1$ leptons, $\leq 4$ jets]
& 1550 GeV & End \\
\hline
%
%
4 & $p_T$(Hardest Lepton) [$\geq 1$ lepton, $\geq 5$ jets] & 150 GeV & End \\
\hline
%
%
5 & $M_{\rm inv}^{\rm jets}$ [0 leptons, $\leq 4$ jets] & 0 GeV & 850 GeV \\
\hline
\end{tabular}\\[2pt]
\end{table*}

To perform the optimization procedure we constructed a large number
of model families in the manner described in
Section~\ref{sec:setup}, each involving the range $-0.5 \leq \alpha
\leq 1.0$ for the parameter $\alpha$ in steps of $\Delta\alpha =
0.05$. For each point along these model lines we generated
100,000~events using {\tt PYTHIA 6.4} and {\tt PGS4} using the
default level-one triggers. To this we added an
appropriately-weighted Standard Model background sample consisting
of 5 fb$^{-1}$ each of $t$/$\bar{t}$ and $b$/$\bar{b}$ pair
production, high-$p_T$ QCD dijet production, single $W^{\pm}$ and
$Z$-boson production, pair production of electroweak gauge bosons
($W^+\,W^-$, $W^{\pm}\,Z$ and $Z\,Z$), and Drell-Yan processes.
Further object-level cuts were then performed,
%
%
followed by
an event-level cut on the surviving detector objects similar to
those used in~\cite{ArkaniHamed:2005px}. Specifically we required
all events to have missing transverse energy $\slashed{E}_T
> 150 \GeV$, transverse sphericity $S_T > 0.1$, and $H_T > 600 \GeV$ (400
GeV for events with 2 or more leptons) where $H_T = \slashed{E}_T +
\sum_{\rm Jets} p^{\rm jet}_T$.


To examine the effectiveness of our candidate signature sets at
measuring the value of the parameter $\alpha$ we fixed ``model~$A$''
to be the point on each of the model lines with $\alpha = 0$ and
then treated each point along the line with $\alpha \neq 0$ as a
candidate ``model~$B$.'' Clearly each model line we investigated --
and each $\alpha$ value along that line -- gave slightly different
sets of maximally effective signatures. The lists we will present
represent an ensemble average over these model lines.

As a straw man we may consider the single most effective signature
at separating models with different values of the parameter
$\alpha$. It is the effective mass formed from all objects in the
event
\begin{equation} M_{\rm eff}^{\rm any} = \met + \sum_{\rm all} p^{\rm
all}_T\, , \label{sigA} \end{equation}
where we form the distribution from all events which pass our
initial cuts. To turn this distribution into a count we integrate
from $M_{\rm eff}^{\rm any} = 1250\GeV$ to the end of the
distribution. We will refer to the variable in~(\ref{sigA}) as
signature ``list''~A. That this one signature would be the most
powerful is not a surprise given the way we have set up the problem.
It is the most inclusive possible signature one can imagine (apart
from the overall event rate itself) and therefore has the largest
overall cross-section.
Furthermore, the variable in~(\ref{sigA}) is sensitive to the mass
differences between the gluino and the lighter electroweak gauginos
-- precisely the quantity that is governed by the parameter
$\alpha$. Yet as we will see in Section~\ref{results} this one
signature can often fail to be effective at all in certain
circumstances, resulting in a rather large required $L_{\rm min}$ to
be able to separate $\alpha =0$ from non-vanishing cases. In
addition it is built from precisely the detector objects that suffer
the most from experimental uncertainty. This suggests a larger and
more varied set of signatures would be preferable.

\begin{table*}[tb]
\caption{Signature List C. Distributions are integrated from ``Min
Value'' to ``Max Value''.} \label{tbl:sigC}
\newcommand{\m}{\hphantom{$-$}}
\newcommand{\cc}[1]{\multicolumn{1}{c}{#1}}
\renewcommand{\tabcolsep}{2pc} 
\renewcommand{\arraystretch}{1.2} 
\begin{tabular}{@{}llcc} \hline
 & Description & Min Value & Max Value \\  \hline
\multicolumn{4}{c}{Counting Signatures} \\ \hline
1 & $N_{\ell}\,\,\quad$ [$\geq 1$ leptons, $\leq 4$ jets] & & \\
2 & $N_{\ell^+ \ell^-}$ [$M^{\ell^+\ell^-}_{\rm inv} = M_Z \pm 5 \GeV$] & & \\
3 & $N_B\,\,\quad$ [$\geq 2$ B-jets] &  &  \\ \hline
\multicolumn{4}{c}{[0 leptons, $\leq 4$ jets]} \\ \hline
4 & $M_{\rm eff}^{\rm any}$ & 1000 GeV & End \\
5 & $M_{\rm inv}^{\rm jets}$ & 750 GeV & End \\
6 & $\met$ & 500 GeV & End \\ \hline
\multicolumn{4}{c}{[0 leptons, $\geq 5$ jets]} \\ \hline
7 & $M_{\rm eff}^{\rm any}$ & 1250 GeV & 3500 GeV \\
8 & $r_{\rm jet}$ [3 jets $>$ 200 GeV] & 0.25 & 1.0 \\
9 & $p_T$(4th Hardest Jet) & 125 GeV & End \\
10 & $\met$/$M_{\rm eff}^{\rm any}$ & 0.0 & 0.25 \\
\hline
\multicolumn{4}{c}{[$\geq 1$ leptons, $\geq 5$ jets]} \\ \hline
11 & $\met$/$M_{\rm eff}^{\rm any}$ & 0.0 & 0.25 \\
12 & $p_T$(Hardest Lepton) & 150 GeV & End \\
13 & $p_T$(4th Hardest Jet) & 125 GeV & End \\
14 & $\met$ + $M_{\rm eff}^{\rm jets}$ & 1250 GeV & End \\
\hline
\end{tabular}
\\[2pt]
\end{table*}

We next consider the five signatures in Table~\ref{tbl:sigB}. These
signatures were chosen by taking our most effective observables and
restricting ourselves to that set for which $\epsilon$ = 10\%. We
again see the totally inclusive effective mass variable
of~(\ref{sigA}) as well as the more traditional effective mass
variable, $M_{\rm eff}^{\rm jets}$, defined via~(\ref{sigA}) but
with the scalar sum of $p_T$ values now running over the jets only.
We now include the $p_T$ of the hardest lepton in events with at
least one lepton and five or more jets, as well as the invariant
mass $M_{\rm inv}^{\rm jets}$ of the jets in events with zero
leptons and 4~or less jets. The various jet-based effective mass
variables would normally be highly correlated with one another if we
were not forming them from disjoint partitions of the overall data
set. The favoring of jet-based observables to those based on leptons
is again largely due to the fact that jet-based signatures will have
larger effective cross-sections for reasonable values of the SUSY
parameters in~(\ref{paramset}) than leptonic signatures. The best
signatures are those which track the narrowing gap between the
gluino mass and the electroweak gauginos and the narrowing gap
between the lightest chargino/second-lightest neutralino mass and
the LSP mass. In this case the first leptonic signature to appear --
the transverse momentum of the leading lepton in events with at
least one lepton -- is an example of just such a signature.

Finally, let us consider the larger ensemble of signatures in
Table~\ref{tbl:sigC}. In this final set we have relaxed our concern
over the issue of correlated signatures, allowing as much as 30\%
correlation between any two signatures in the list. This allows for
a larger number as well as a wider variety of observables to be
included. This can be very important in some cases in which the
supersymmetric model has unusual properties, or in cases where the
two $\alpha$ values being considered give rise to different mass
orderings (or hierarchies) in the superpartner spectrum. In
displaying the signatures in Table~\ref{tbl:sigC} we find it
convenient to group them according to the partition of the data
being considered. Note that the counting signatures are taken over
the entire data set.

The first counting signature is simply the total size of the
partition in which the events have at least one lepton and 4~or less
jets. The next two signatures are related to ``spoiler'' modes for
the trilepton signal. Note that the trilepton signal itself did {\em
not} make the list: this is a wonderful discovery mode for
supersymmetry, but the event rates between a model with $\alpha =0$
and one with non-vanishing $\alpha$ were always very similar (and
low). This made the trilepton counting signature ineffective at
distinguishing between models. By contrast, counting the number of
b-jet pairs (a proxy for counting on-shell Higgs bosons) or the
number of opposite-sign electron or muon pairs whose invariant mass
was within 5~GeV of the Z-mass (a proxy for counting on-shell
Z-bosons) {\em were} excellent signatures for separating models from
time to time. This was especially true when the two models in
question had very different values of $\alpha$ such that the mass
differences between $\wtd{N}_2$ and $\wtd{N}_1$ were quite different
in the two cases. Signature \#8 is defined as the following ratio
\begin{equation} r_{\rm jet} \equiv \frac{p_T^{\rm jet 3} + p_T^{\rm
jet 4}}{p_T^{\rm jet 1} + p_T^{\rm jet 2}} \label{rjet}
\end{equation}
where $p_T^{\rm jet\,i}$ is the transverse momentum of the $i$-th
hardest jet in the event. For this signature we require that there
be at least three jets with $p_T > 200 \GeV$. This signature, like
the $p_T$ of the hardest lepton or the $p_T$ of the 4th~hardest jet,
was effective at capturing the increasing softness of the products
of cascade decays as the value of $\alpha$ was increased away from
$\alpha = 0$.

\section{RESULTS}
\label{results}

In this section we examine how well our signature lists in perform
in measuring the value of the parameter $\alpha$ which appears
in~(\ref{mirage_ratios}). These lists were designed through the
analysis of several hundred ``model lines'' as described previously.
To test the approach we generated a random collection of 500 new
models with six points on the $\alpha$-lines ranging from 0 to 0.5.
In this case a~16-dimensional parameter space defined by the
quantities in~(\ref{paramset}) was considered. Specifically, slepton
and squark masses were allowed to vary in the range 300~GeV to
1200~GeV with the masses of the first and second generation scalars
kept equal. The gaugino mass scale given by $M_3$ and the
$\mu$-parameter were also allowed to vary in this range. The
pseudoscalar Higgs mass $m_A$ was fixed to be 850 GeV and the value
of $\tan\beta$ was allowed to vary from 2~to~50. If all points along
the $\alpha$-line satisfied all experimental constraints on the
superpartner mass spectrum, then 100,000 events were generated for
each of the six points along the $\alpha$-line in the manner
described in Section~\ref{sec:simulation}. Using this data the value
of $L_{\rm min}$ was computing using~(\ref{Lmin}) for each of our
three signature sets.

\begin{figure}[t]
\begin{center}
      \includegraphics[scale=0.4]{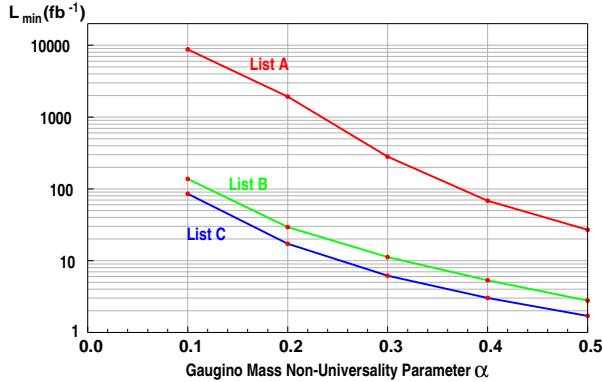}
\caption{\label{fig:results1}\footnotesize{\textbf{$L_{\rm min}$
required to detect $\alpha \neq 0$ for 95\% of the random models.}
}}
\end{center}
\end{figure}

\begin{figure}[t]
\begin{center}
      \includegraphics[scale=0.4]{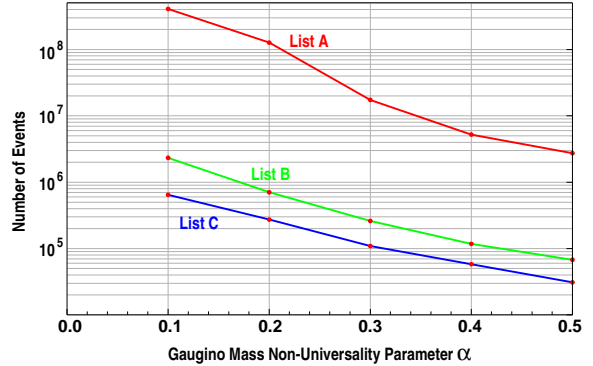}
\caption{\label{fig:results2}\footnotesize{\textbf{$N_{\rm min}$
required to detect $\alpha \neq 0$ for 95\% of the random models.}
}}
\end{center}
\end{figure}

The results of this analysis are given in Figures~\ref{fig:results1}
and~\ref{fig:results2}. In Figure~\ref{fig:results1} the integrated
luminosity needed to detect $\alpha \neq 0$ for 95\% of our random
models is given as a function of the five non-vanishing $\alpha$
values simulated. Since the random model sample includes examples
with very different superpartner mass scales, the overall
supersymmetric production cross-section varies much more across this
sample than in the controlled model sample described above. We
therefore take this into account by plotting the required number of
supersymmetric events in Figure~\ref{fig:results2}. While our single
best signature (``List A'') has some resolving power, it is the
larger sets that will prove most effective in establish the
non-universality of gaugino masses, as to be expected. Yet going
from the 5~signatures of List~B to the 14~signatures of List~C
produces only a marginal increase in resolving power. This is
evidence of the increased correlations amongst the signatures of
List~C. It is also demonstrative of the general principle of
diminishing returns to adding new observables to any particular set.
Even using our best set of signatures (List~C) it will require
nearly 100~${\rm fb}^{-1}$ to be able to detect non-universality at
the level of $\alpha \simeq 0.1$ for an arbitrary supersymmetric
model. Yet for the vast majority of models the departure from
universality should become apparent after just 10-20~${\rm
fb}^{-1}$. Departures from universality at the level of $\alpha
\simeq 0.3$ should be apparent using this method for most
supersymmetric models after just a few~${\rm fb}^{-1}$.

\begin{figure}[p]
\begin{center}
      \includegraphics[scale=0.5]{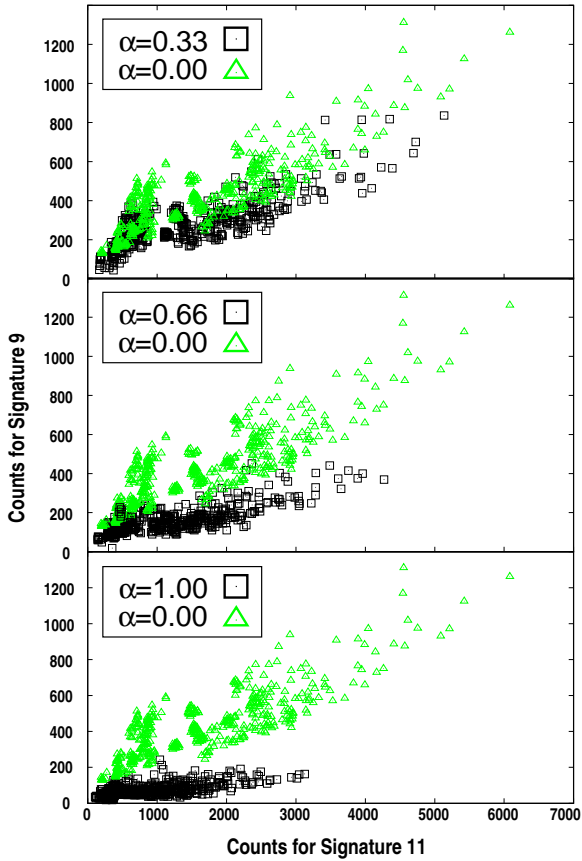}
\caption{\label{fig:scatterC}\footnotesize{\textbf{Footprint-style
plot for a pair of signatures from List~C.} Total counts for
signature \#11 versus signature \#13 of List~C is given for the case
$\alpha = 0$ (green triangles) $\alpha \neq 0$ (black squares). The
cases shown are for $\alpha = 0$ versus $\alpha = 0.33$ (top panel),
$\alpha = 0.66$ (middle panel) and $\alpha = 1$ (bottom panel). The
axes measure the number of events for which the kinematic quantity
was in the range given in Table~\ref{tbl:sigC}. Larger values of the
non-universality parameter $\alpha$ correspond to a greater degree
of separation between the two model ``footprints.''}}
\end{center}
\end{figure}

To understand why this approach works, it is useful to examine the
signature results themselves. As an example, consider a pair of
signatures drawn from List~C in Table~\ref{tbl:sigC}.
Figure~\ref{fig:scatterC} shows a two-dimensional slice of the
signature space ``footprint'' for our large set of model
variations~\cite{Bourjaily:2005ja,Kane:2006yi,Kane:2007pp}. In these
figures the results have been normalized to $5\;{\rm fb}^{-1}$ of
data. The count rate for signature \#9 is shown versus that of
signature \#11 of List~C for the case $\alpha = 0$ and $\alpha =
0.33$ (top panel), $\alpha = 0.66$ (middle panel) and $\alpha = 1$
(bottom panel).
%
%
We have chosen this pair for the dramatic separation that can be
achieved, though similar results can be obtained with other pairs of
signatures.

The power of our inclusive signature list approach lies in the
choice of signatures and their ability to remain highly sensitive to
changes in the physical behavior of each model. This feature is
reflected qualitatively in the visual clustering of the data points,
which become progressively more distinct as the parameter~$\alpha$
is increased. As the regions separate it becomes increasingly less
likely that a model from one class can be confused with a model from
the other class, even when considering statistical fluctuations. In
our approach this manifests itself when one computes $R_{AB}$, which
reflects the ``distance'' in signature space between the two models
under comparison, and which becomes large when the models are
sufficiently different from one another.

\section{CONCLUSIONS}

If supersymmetry is discovered at the LHC the high energy community
will be blessed with a large number of new superpartners whose
masses and interactions will need to be measured. Undoubtedly
performing global fits of the many observables to the parameter
space of certain privileged and well-defined benchmark models will
be of great help in making sense of this embarrassment of richness.
It is an interesting question to ask whether it is possible to fit
to certain model {\em characteristics} rather than to any particular
model itself.
Perhaps the most important such characteristic is the pattern of
soft supersymmetry-breaking gaugino masses. No other property of the
low-energy soft Lagrangian is more easily linked to underlying
high-scale physics, particularly if that high-scale physics is of a
string-theoretic origin. We are thus interested in asking whether we
can identify the presence on non-universalities in the gaugino
sector independent of all other properties of the superpartner
spectrum. In the present work we have decided to begin this process
with a simple parametrization in terms of model ``lines'', in the
spirit of previous benchmark studies such as the Snowmass Points \&
Slopes~\cite{Allanach:2002nj} in which only the single
non-universality parameter is varied. By understanding how the
observable physics at the LHC is affected by this parameter -- and
then repeating the analysis many times with the other supersymmetric
parameters varied -- we were able to obtain two sets of observables
that fared well in detecting the presence of non-universalities with
relatively small amounts of integrated luminosity.
The larger of the two sets generally performed slightly better, but
at the expense of allowing signatures that are correlated with one
another at the 30\% level. Requiring a correlation at only the 10\%
level (and thus using a shorter list of observables) had only a
small effect on the ability to distinguish the size of the
non-universality parameter $\alpha$.
Broadly speaking, we find that a non-universality at the 10\% level
can be measured with 10-20 fb$^{-1}$ of integrated luminosity over
approximately 80\% of the supersymmetric parameter space relevant
for LHC observables. If we are interested in measurements at only
the 30\% level these numbers change to 5-10 fb$^{-1}$ over
approximately 95\% of the relevant parameter space.

\end{document}